\documentclass[]{elsart}
\usepackage{graphicx}
\usepackage{amssymb}
\newcommand{\kam}{Kamioka Observatory}
 
\begin{document}
\intextsep 30pt
\textfloatsep 30pt
\begin{frontmatter}
\title{Detecting the WIMP-wind via spin-dependent interactions}
\author{Toru~Tanimori},
\author{Hidetoshi~Kubo},
\author{\corauthref{cor1}Kentaro~Miuchi},
\ead{miuchi@cr.scphys.kyoto-u.ac.jp}
\author{Tsutomu~Nagayoshi},
\author{Reiko~Orito},
\author{Atsushi~Takada},
\author{Atsushi~Takeda} 
\address{Cosmic-Ray Group, Department of Physics, Graduate School of Science, 
Kyoto University Kitashirakawa, Sakyo-ku, Kyoto, 606-8502, Japan}
\corauth[cor1]{Corresponding author.}
\begin{abstract}
Revealing the nature of dark matter is one of the most interesting 
tasks in astrophysics. 
Measuring the distribution of recoil angles is said to be 
one of the most reliable methods to detect a positive 
signature of dark matter.
We focused on measurements via spin-dependent interactions, 
and studied the feasibility with carbon tetrafluoride($\rm CF_4$) gas, 
while taking into account the performance of 
an existing three-dimensional tracking detector. 
We consequently found that it is highly possible to detect a 
positive signature of 
dark matter via spin-dependent interactions.
\end{abstract}

\end{frontmatter}
\newpage
\section{Introduction}
\label{section:intro}
Existence of the dominant Cold Dark Matter(CDM) became
much more concrete by the recent results of the WMAP 
cosmic microwave background(CMB) 
all-sky observation with other finer scale CMB measurements (ACBAR and CBI), 
2dFGRS measurements, and Lyman $\alpha$ forest data \cite{ref:WMAP Spergel}.
Best fit cosmological parameters, $h=0.71^{+0.04}_{-0.03}$, 
$\Omega _{\rm b}h^2=0.0224\pm 0.0009$, 
$\Omega_{\rm m} h^2=0.135^{+0.008}_{-0.009}$, and 
$\Omega_{\rm tot} h^2=1.02\pm 0.02$, 
show that the CDM consists about 20$\%$ of the energy
and dominates the mass of the universe.

Weakly Interacting Massive Particles (WIMPs) 
are one of the best candidates for CDM.
In spite of quite a few WIMP search experiments, 
there has not yet been any experimental 
evidence of WIMP detection \cite{ref:Nelson_review}, 
except for an indication of an annual modulation signal, 
reported by the DAMA group \cite{ref:DAMA annual}.
Because the amplitude of an annual modulation signal is 
only a few $\%$, 
positive signatures of WIMPs are 
very difficult to detect 
with conventional methods, which basically measure the recoil spectrum.
Owing to the motion of the solar system with respect to the galactic halo,
the direction-distribution of the WIMP velocity that we observe 
at the earth is expected to show an asymmetry, like a wind of WIMPs. 
Attempts to detect a positive signature of WIMPs
by measuring the recoil angles have been carried 
out \cite{ref:DAMA_aniso,ref:Buckland_PRL,ref:APP_CH4,DRIFT_NIM,DRIFT_IDM,ref:Pikachu} 
ever since it was indicated to be an alternative and more reliable 
method \cite{ref:Spergel_WIMP}.
Gaseous detectors are one of the most appropriate 
devices for detecting this WIMP-wind\cite{ref:Gerbier,ref:Masek}. 
Properties of 
the $\rm CS_2$ gas, which is sensitive to 
the WIMP-wind mainly via 
spin-independent (SI) interactions are 
mainly studied because of its small diffusions\cite{DRIFT_NIM,DRIFT_IDM}.
We, on the other hand, are focusing on the detection of WIMPs via 
spin-dependent (SD) interactions, because WIMPs can basically be detected 
both via SI and SD interactions.
Because fluorine was said and also found to be very effective for a 
WIMP search via SD interactions, 
as shown in previous works \cite{ref:Ellis91,ref:SIMPLE,ref:miuchi_APP}, 
we studied the detection feasibility 
with carbon tetrafluoride ($\rm CF_4$) gas,
which has been studied very well as 
one of the standard gases for 
time projection chambers (TPC) \cite{ref:NIM_CF4_Schmidt,ref:NUMU}.
While the use of $\rm CF_4$ has been proposed by others 
before\cite{ref:NUMU_DM,ref:Collar_NIM}
, this is the first in-detail feasibility study,
taking into account the performance of an existing 
three-dimensional tracking detector ($\mu$-TPC) and the measured 
neutron background flux.

\section{$\mu$-TPC}
\label{section:micto-TPC}
\begin{figure}[h]
   \begin{center}
\includegraphics[width=0.9\linewidth]{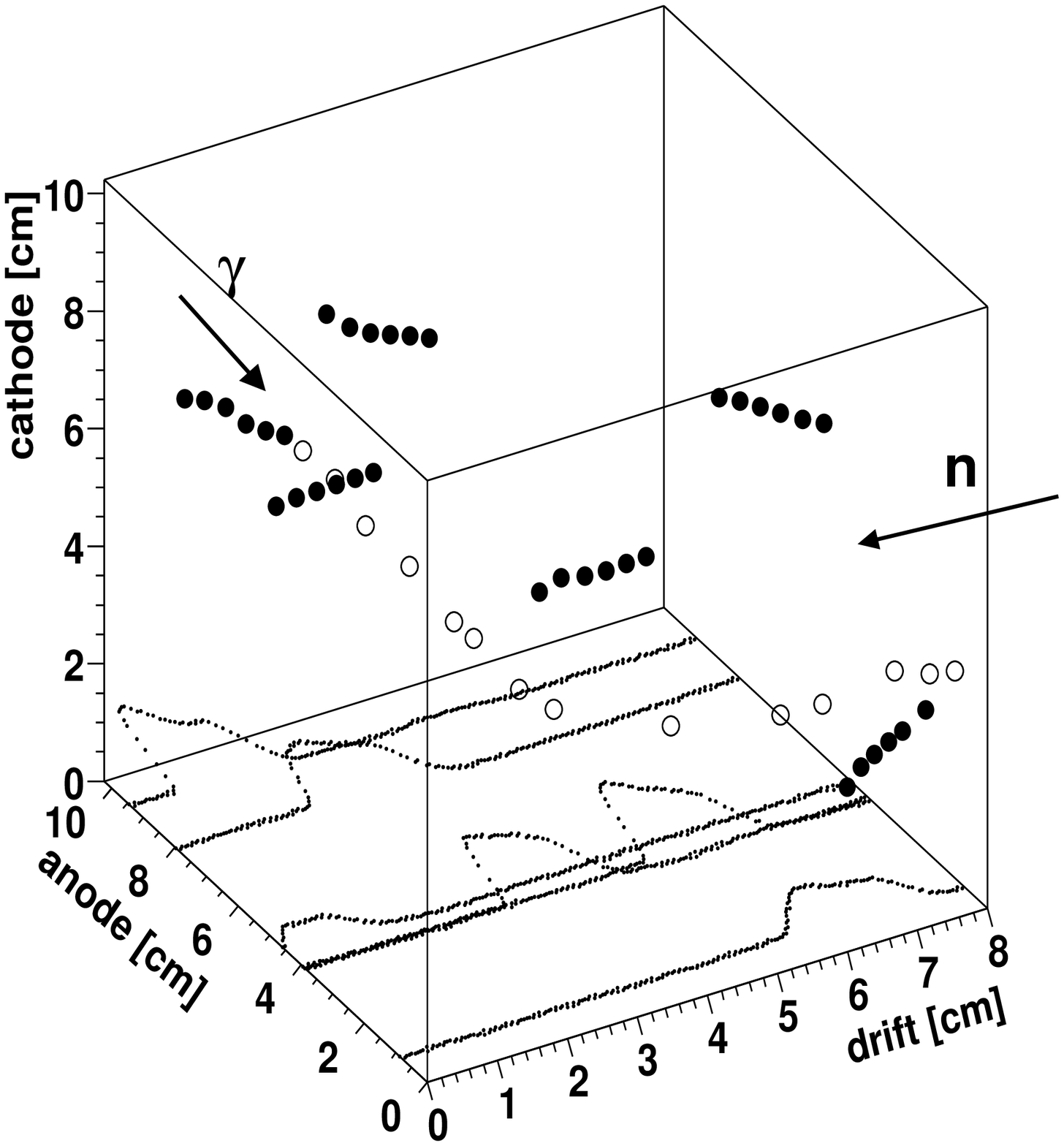}
   \caption{Several three-dimensional proton tracks 
and their Bragg curves detected by the $\mu$-TPC filled 
with a Ar-$\rm C_2 H_6$ gas mixture.
The $\mu$-TPC was irradiated with the 
fast neutrons and gamma-rays from a radioactive source of  
$\rm {}^{252}Cf$. 
Detected tracks of the recoil protons (500 keV - 1 MeV) are shown by the 
filled circles.
The FADC data, which can be recognized as the Bragg curves are shown in the 
cathode=0cm plane.
A typical track of a Compton-scattered electron 
is shown by blank circles for reference.
The arrows indicate the directions of the incoming primary particles.
}

\label{fig:track}
  \end{center}
   \end{figure}

A $\mu$-TPC is a time projection chamber with a micro pixel chamber 
($\mu$-PIC) readout, being developed for the detection of tracks 
of charged particles with fine spatial resolutions
 \cite{uPIC:Ochi,ref:Nagayoshi_PSD,TPC_PSD:Kubo,IEEE:Miuchi}. 
A $\mu$-PIC is a gaseous two-dimensional position-sensitive detector 
manufactured using printed circuit board (PCB) technology. 
With PCB technology, large-area detectors can, in principle, be 
mass-produced, 
which is an inevitable feature for a dark matter detector.
We developed a prototype $\mu$-TPC with a detection volume of 
10 $\times$ 10 $\times$ 8 $\rm cm^3$, and 
studied its fundamental properties 
with a Ar-$\rm C_2 H_6$ gas mixture of 1atm. 
Anode electrodes are formed with a pitch of 400 $\mu$m, and  
signals from anode strips and cathode strips are amplified 
and discriminated into LVDS-level signals. 
Discriminated digital signals are fed to an 
FPGA-based position encoding module and 
synchronized with an internal clock of 20 MHz,
so that two-dimensional hit positions can be calculated by the 
anode-cathode coincidence within one clock. 
Together with the 
timing information, three-dimensional tracks are thus detected as successive 
points. 
Measured two-dimensional ($\mu$-PIC intrinsic) and three-dimensional position 
resolutions of the $\mu$-TPC are 120$\mu$m\cite{ref:takeda_IEEE} 
and 260$\mu$m\cite{ref:nagayoshi_imaging2003}, respectively.
We irradiated the $\mu$-TPC with fast neutrons from a radioactive source of 
$\rm {}^{252}Cf$ \cite{ref:Miuchi neutron}, and 
tracks of the recoil protons (500 keV - 1 MeV) were detected, as shown 
by the filled circles in FIG. \ref{fig:track}. 
The distance between the detected points was 
restricted by the clock of the 
electronics (20 MHz), which we will soon increase to more than 50 MHz
\cite{ref:nagayoshi_imaging2003}. 
This result indicates that a $\mu$-TPC, which can detect 
tracks of the charged particles down to 3 mm, will be
available for the dark matter search experiment.
The sums of the all cathode signals recorded by a flash ADC (FADC)
are also shown in FIG.~\ref{fig:track} 
in the cathode=0 plane. 
We recognized that 
Bragg curves were well-reproduced by the FADC data, 
and thus the track directions were obviously known.
In FIG. \ref{fig:track},  a typical electron track of about 100 keV  
is also shown by the blank circles.
Because the electron tracks are much more winding, and have a smaller 
energy deposition ($dE/dx$) than those of  
protons, gamma-rays are known to be discriminated 
with high efficiency \cite{ref:Miuchi neutron}. 
Stable operation of the $\mu$-TPC with a 
$dE/dx$ threshold of 15~keV/cm (gain $\sim$ 5000) 
for more than 1000 hours was already realized. 
We are improving the geometrical structure of the 
electrodes by using three-dimensional 
simulators \cite{ref:Finland}, which we expect will
increase the gas gain by a factor of three.
Then, the $dE/dx$ threshold will be better than
10 keV/cm, even with $\rm CF_4$ gas, 
which has a larger $W$ value (average energy to produce one electron-ion pair) 
than argon($W_{\rm CF_4}$=54 eV, $W_{\rm Ar}$=26 eV\cite{ref:Sharma}).
We started a 
study of the detector performance with $\rm CF_4$ gas, and 
found that $\mu$-TPC shows a good detector performance 
(gas gain $\sim$~10000).
Precise studies on the gas properties will be reported in another paper.
Based on the detector studies described above, and 
on-going improvements,
we assume that a $\mu$-TPC as 
a dark matter detector have the properties listed below:
\begin{itemize}
\item Track length threshold is 3 mm.
\item $dE/dx$ threshold is 10 keV/cm.
\item Gamma-ray discrimination efficiency is close to 100$\%$.
\item Bragg curves are detected.
\end{itemize}

\section{WIMP-wind measurement}
\label{section:calculation}
\begin{figure}[h]
   \begin{center}
\includegraphics[width=1.0\linewidth]{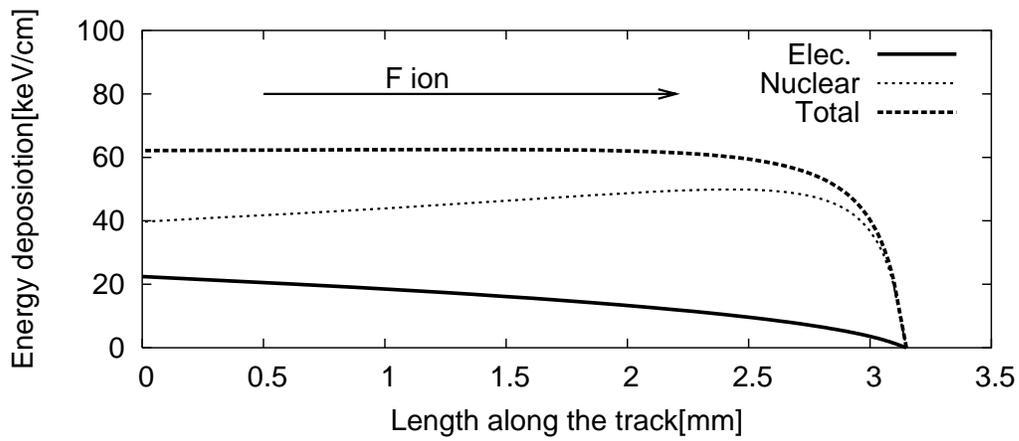}
   \caption{Calculated energy loss of a F ion of 25~keV in 20~Torr $\rm CF_4$ gas. The energy loss in the electron field, nuclear filed, and the total energy loss are shown by the solid, dotted, and dashed lines, respectively.}
 \label{fig:F_Bragg}
  \end{center}
   \end{figure}

We studied the feasibility of WIMP-wind detection with $\rm CF_4$ gas 
considering the detector characteristics listed above.
First, we calculated the energy deposition of a F 
ion in $\rm CF_4$ gas. 
We set the gas pressure at 20 Torr and the recoil energy at 25~keV. 
Because the binding energy of a C-F in a $\rm CF_4$ molecule is 4.6 eV, 
all of the recoil energy is thought to become 
kinetic energy of the F ion.
The energy deposition was calculated by SRIM2003 \cite{ref:SRIM} and the 
range was scaled by the same manner as discussed in Ref. \cite{DRIFT_NIM}.
The energy deposition was also scaled by the same scaling factor.
The range $R$ of the F ion in $\rm CF_4$ is expressed 
as a function of the pressure $p$ and the recoil energy $E$ 
with a good approximation down to 
$E$=5~keV in this situation as follows:
\begin{equation}
R=\left(\frac{dE}{dx}\right)_{\rm Ne_{in}Ar}^{-1}\frac{N_{\rm Ar}}{N_{\rm CF_4}}E  \frac{\rm 760 Torr}{p}=k \cdot E {p}^{-1},
\end{equation}
where $\left(\frac{dE}{dx}\right)_{\rm Ne_{in}Ar}$ 
$\sim$~130keV/mm \cite{ref:Gilbert} is the energy deposition 
of a Ne ion in Ar gas;  
$\frac{N_{\rm Ar}}{N_{\rm CF_4}}$ is the ratio of the electron density of 
Ar gas and $\rm CF_4$ gas.
We consequently knew that a 25~keV F ion has a range of 
roughly 3 mm in 20 Torr 
$\rm CF_4$ using k=2.63 $\rm mm\cdot Torr/keV$.
The calculated energy losses of a F ion in 
$\rm CF_4$ gas are shown in FIG. \ref{fig:F_Bragg}. 
The range of an ion is determined by the 
total (electron + nuclear fields) energy loss, shown in the dashed line. 
Since the energy loss in the nuclear field has 
a negligible effect on ionizations, 
the ''visible'' energy loss is that in the electron 
field shown by the solid line.
From the calculated energy loss in the electron field, it is seen that 
the  electron diffusion should be suppressed to lower than 
$\sigma <$ 1 mm in order to know the track direction.

\begin{figure}[h]
   \begin{center}
\includegraphics[width=1.0\linewidth]{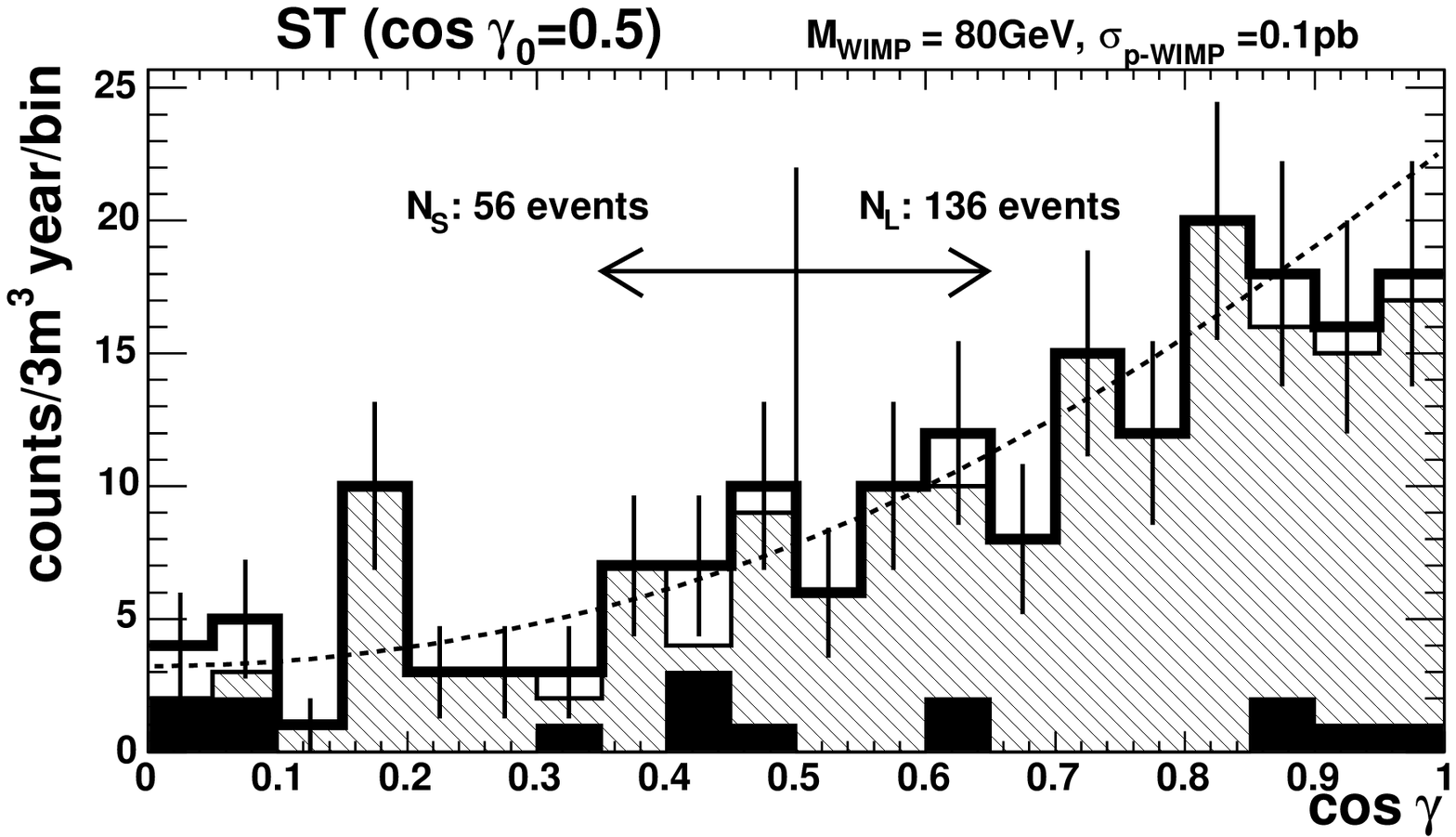}
\includegraphics[width=1.0\linewidth]{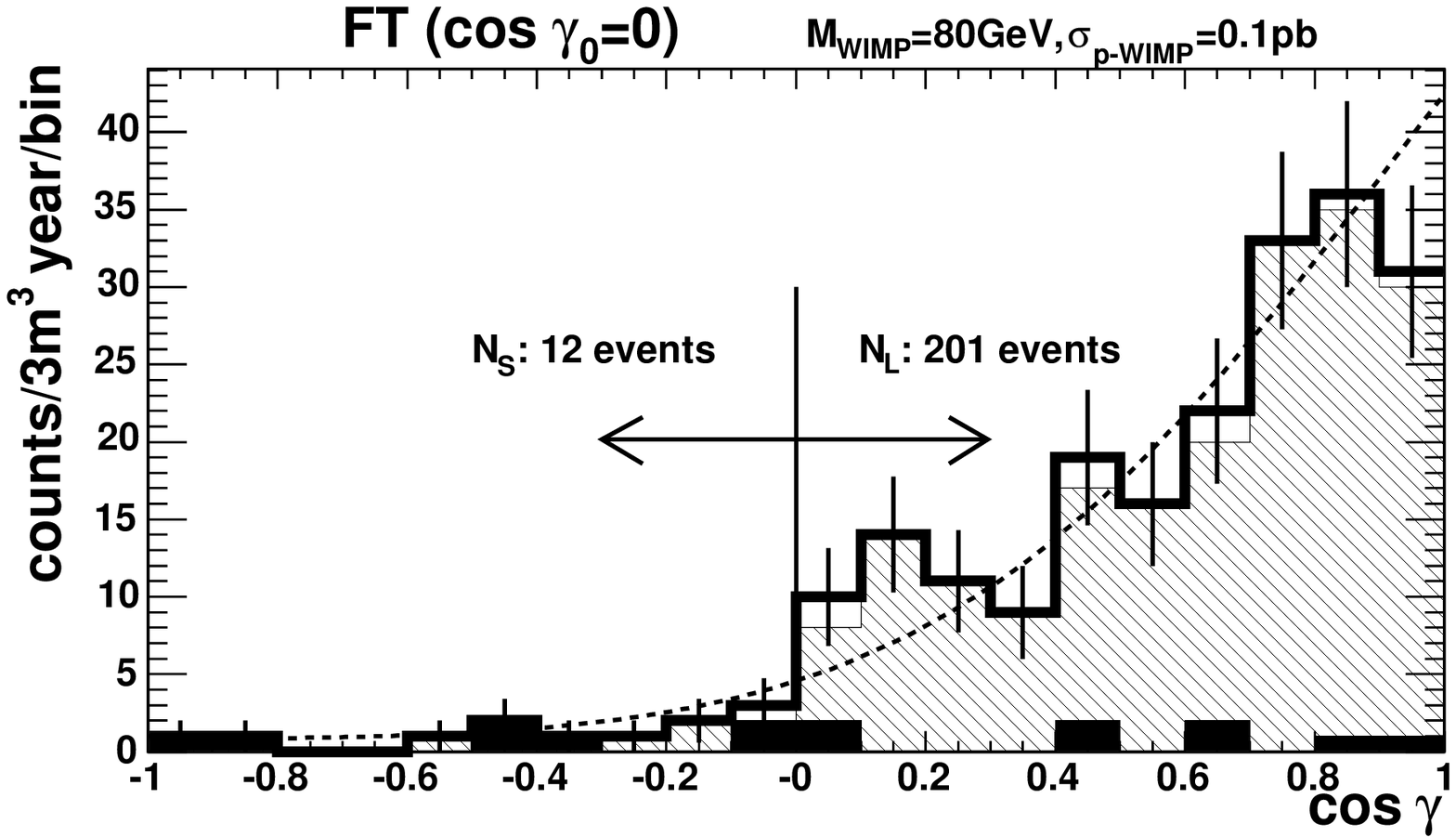}
   \caption{Simulated recoil angle distributions for the ST(semi-tracking) mode (upper) and the FT(full-tracking) mode(lower). The pressures of the 
$\rm CF_4 $ gas are 30 Torr 
and 20 Torr for the ST mode and FT mode, respectively.
Neutron background is estimated based on the 
measured fast neutron flux((1.9 $\pm$ 0.21) $\rm \times 10^{-6} n\cdot cm^{-2}\cdot s^{-1}$) at {\kam} and a 50cm water shield. 
WIMP signals, neutron background, and the total observable events are shown in the hatched, filled, and solid-line histograms, respectively. Theoretical calculations are also shown by the dashed lines. }
 \label{fig:F_asyn}
  \end{center}
   \end{figure}

In the following discussion, we consider two experimental modes:
\begin{enumerate} 
\item We use the digital hit points for tracking, but do not use the 
FADC data for determining the track direction.
In this case, the recoil angle is known only by the 
absolute value of its cosine.
We hereafter call this semi-tracking, or the ''ST'' mode. 
\item We somehow suppress the electron diffusion and use the full information 
on their tracks, including the direction. 
We hereafter call this full-tracking, or the ''FT'' mode. 
\end{enumerate} 
The FT mode will be realized by limiting the drift length, 
or by using an elencro-negative gas \cite{ref:DRIFT_CS2}. 
A magnetic field doesn't help, however, because 
the longitudinal electron diffusion cannot be suppressed by the magnetic 
field. 
The electron diffusion ($\sigma$) along the track path is expressed 
as $\sigma=\sqrt{2DL/W}$, where D is the diffusion coefficient, $L$ is the 
drift length, and $W$ is the drift velocity.
$D<$ 1000 $\rm cm^2/s$ \cite{ref:NIM_CF4_Schmidt} and $W>$ 10 cm/$\mu$s 
are achieved for a 
reduced electric field of $E/p>1$kV/cm/atm, 
which implies that $\sigma <$ 1 mm is realized with 
a drift length of 50 cm in 20 Torr of $\rm CF_4$ gas. 
We calculated the recoil angle distributions for the ST and FT modes.  
We followed Ref.  \cite{ref:LewinSmith} for the energy-spectrum 
calculation and
\begin{equation}
\frac{dR}{dEd\cos \gamma} \propto \exp\left[ \frac{ \left( v_{\rm s} \cos \gamma -v_{\rm min}\right) ^2}{v^{2}_{0}} \right]
\end{equation}
in Ref.  \cite{ref:Buckland_PRL,ref:Spergel_WIMP} 
for the direction distribution calculation.
Here, $R$ is the count rate, $E$ is the recoil energy, 
$\gamma$ is the recoil angle,
$v_{\rm s}$ is the solar velocity with respect to the galaxy, 
$v_{\rm min}$ is the minimum velocity of 
WIMPs that can give a recoil energy of $E$,
and $v_{0}$ is the Maxwellian WIMP velocity dispersion.
We used the astrophysical, nuclear, and experimental 
parameters given in Table \ref{table:app and nu params}. 
We assumed that the gamma-rays, $\beta$-rays, and $\alpha$-rays 
are discriminated 
by 100$\%$ efficiency, and that fast neutrons dominate the background.
This assumption is reasonable because we can detect   
three-dimensional tracks with fine samplings 
in contrast to the two-dimensional 
tracks discussed in Ref. \cite{DRIFT_NIM}.
The measured fast neutron flux at {\kam} \cite{ref:n_at_Kam} 
was used for the calculation. 
\begin{table}[h]
\begin{center}
\begin{tabular}{l l}
\hline
WIMP velocity distribution & Maxwellian  \\ \hline
solar velocity & $v_{\rm s}$=244km$\cdot \rm {s}^{-1}$ \\ \hline
Maxwellian velocity dispersion& $v_{0}$=220km$\cdot \rm {s}^{-1}$ \\ \hline
escape velocity &$v_{\rm esc}$ =650 km$\cdot \rm s^{-1}$ \\ \hline
local halo density & $\rm 0.3\,GeV\cdot cm^{-3}$ \\ \hline
spin factor of $\rm {}^{19}F$ & $\lambda^2J(J+1)=0.647$ \\ \hline \hline
target gas & $\rm CF_4 $ \\ \hline
gas pressure (ST/FT)& 30 Torr / 20 Torr \\ \hline
energy threshold (ST/FT)& 35 keV / 25 keV\\ \hline
track length threshold & 3 mm \\ \hline
fast neutron flux & (1.9 $\pm$ 0.21) $\rm \times 10^{-6} n\cdot cm^{-2}\cdot s^{-1}$ \\ \hline
neutron shield & 50 cm water \\ \hline \hline
$M_{\rm WIMP}$ &80 GeV \\ \hline
$\sigma^{\rm SD}_{\rm WIMP-p}$ & 0.1pb \\ \hline
exposure & 3 $\rm m^3\cdot year$  \\ \hline
\end{tabular}
\caption{Astrophysical, nuclear, and experimental parameters used for the calculations. The density of 30 Torr of $\rm CF_4 $ is 155 $\rm g\cdot m^{-3}$}
\label{table:app and nu params}
\end{center}
\end{table}
The simulated $\cos \gamma$ distributions for 
the ST and FT modes are shown in the upper and lower panels of 
FIG. \ref{fig:F_asyn}, respectively.
The WIMP signals are shown by the hatched histograms,
the neutron backgrounds are shown in the filled histograms, 
and the total observable events are shown in the histograms with errorbars.
We simply analyzed the asymmetry by comparing the number of 
events with $\cos \gamma > \cos \gamma_0$ ($N_{\rm L}$) and that with 
$\cos \gamma < \cos \gamma_0 $ ($N_{\rm S}$), 
where $\gamma_{0}$ was chosen so that $N_{\rm L}$ equals $N_{\rm S}$
for a flat background. 
The asymmetry-detection confidence level($ADCL$) was then defined as
\begin{equation}
ADCL=\frac{|N_{\rm L}-N_{\rm S}|}{\sqrt{N_{\rm L}+N_{\rm S}}}.
\end{equation}
We obtained 5.8 $\sigma$ and 13 $\sigma$ in the ST mode 
and the FT mode, respectively.
Although the ST mode showed a lower confidence level than the FT mode, 
it was shown that the ST mode is still a very strong 
method compared to the conventional method
to detect the positive signal of WIMPs.
Because the direction of the WIMP-wind has a diurnal modulation which 
goes out of phase with a period of a year,
it is unlikely 
that any background that is isotropically distributed or 
localized near the detector, would mimic this asymmetry.
Event the background that have or could have fixed day/night 
circadian rhythm does not mimic the WIMP signal because WIMP-wind 
is expected to be blowing in 
the same direction during the day now and during the night in six months.

\begin{figure}[h]
   \begin{center}
\includegraphics[width=1.0\linewidth]{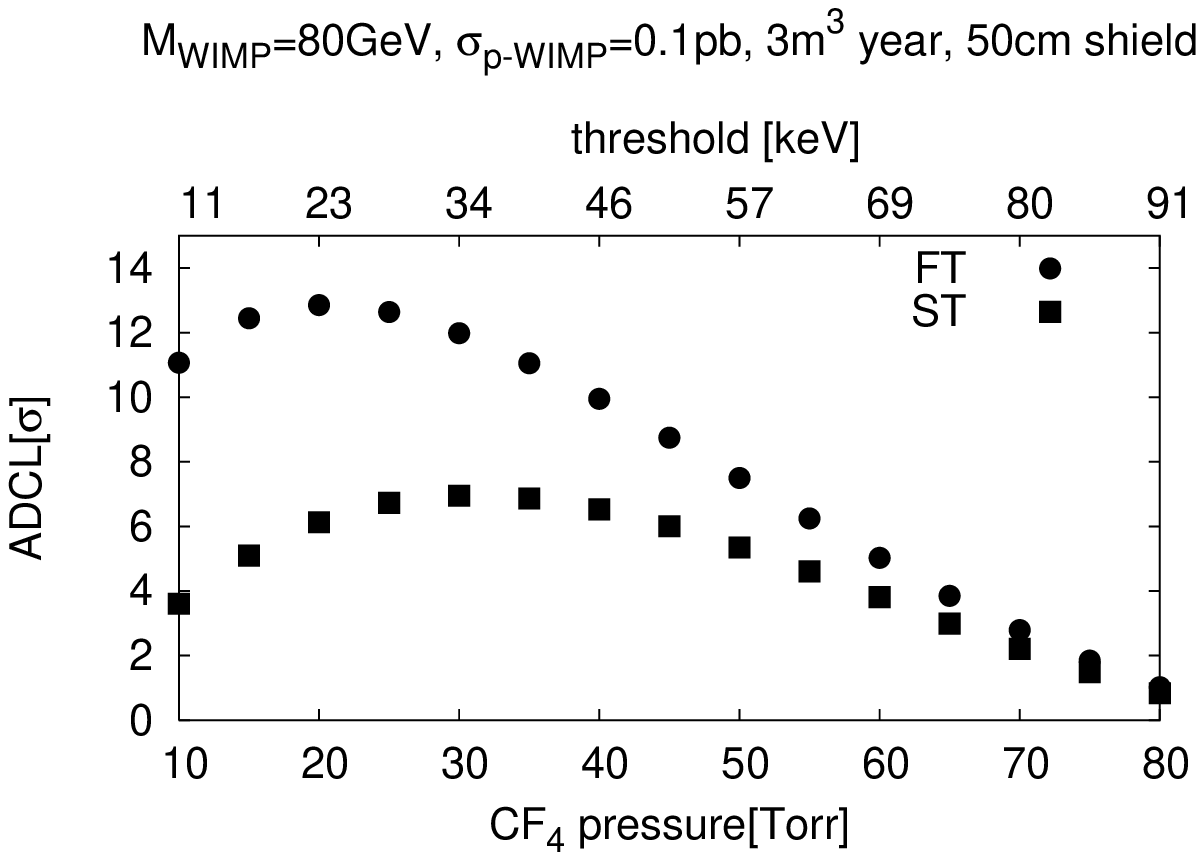}
   \caption{Asymmetry-detection confidence level($ADCL$) [$\sigma$] 
as a function of the gas pressure and the corresponding 
recoil energy threshold. 
It is seen that FT(full-tracking) and ST(semi-tracking) 
modes have different optimum pressures.}
 \label{fig:p-t_opt}
  \end{center}
   \end{figure}

\begin{figure}[h]
   \begin{center}
        \includegraphics[width=1.0\linewidth]{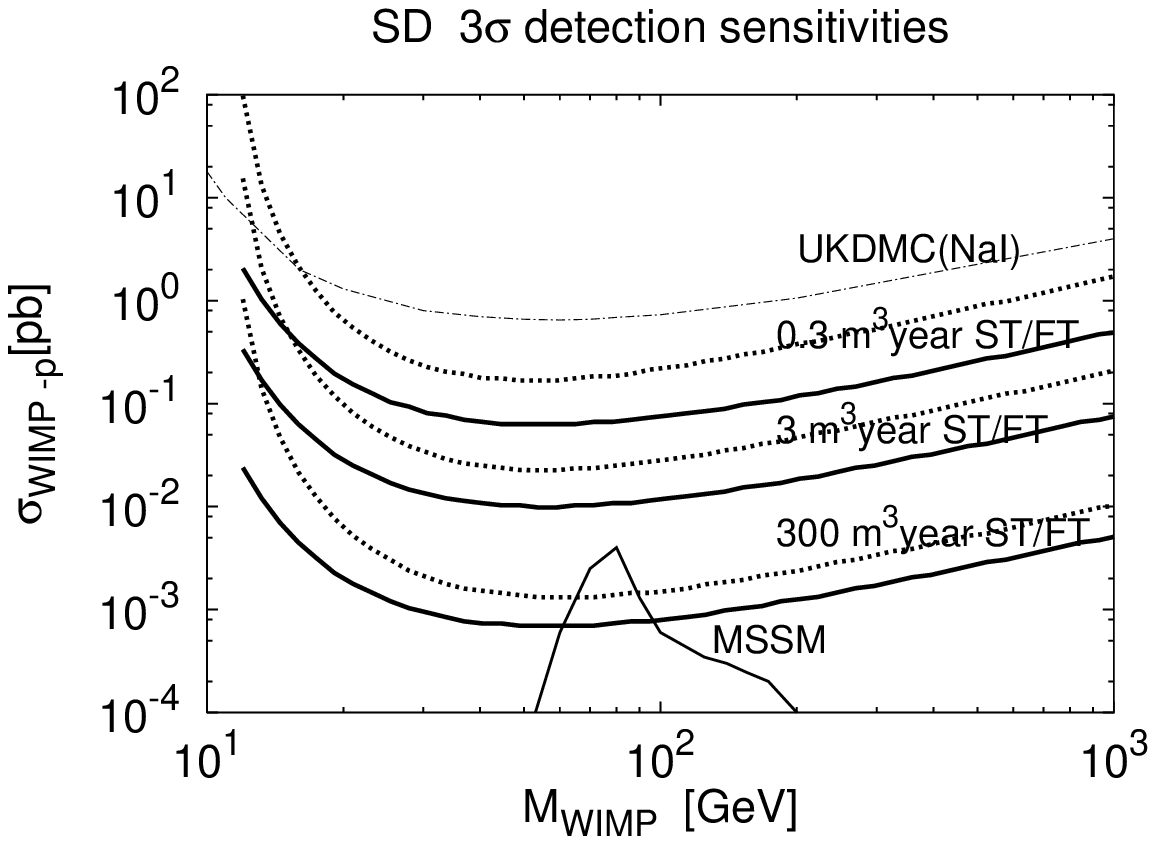}
   \caption{Simulated 3$\sigma$ asymmetry detection sensitivities 
by the measurement at {\kam}.
Sensitivities by the ST(semi-tracking) and FT(full-tracking) 
modes are shown by the thick dotted and thick solid lines, respectively.
Three exposures are considered for each mode.
The pressures of the 
$\rm CF_4 $ gas are 30 Torr 
and 20 Torr for the ST mode and FT mode, respectively.
An experimental result of UKDMC (dashed-dotted) and MSSM predictions (solid) are also shown \cite{ref:uk2000,SUSY_Ellis}.}
\label{fig:limit}
  \end{center}
   \end{figure}

We calculated $ADCL$ values for several  
parameter sets in order to evaluate 
its dependence on the experimental parameters.
The experimental parameters given in the TABLE \ref{table:app and nu params}, 
except for the gas pressure and the energy threshold
are used for the calculation. The result 
is shown in FIG. \ref{fig:p-t_opt}.
It is seen that 
the ST mode and the FT mode have 
the largest $ADCL$ values at 30 Torr and  20 Torr, respectively.
In FIG.\ref{fig:limit}, 3 $\sigma$ detection sensitivities to the 
$\sigma^{\rm SD}_{\rm WIMP-p}$
as a function of $M_{\rm WIMP}$ are shown.
The parameters given in TABLE \ref{table:app and nu params}, except for 
$M_{\rm WIMP}$, $\sigma^{\rm SD}_{\rm WIMP-p}$, and the exposure, were used 
for the calculation.
Here, the 3 $\sigma$ detection level was defined as the smallest cross section 
for which we observe $ADCL$ = 3~$\sigma$.
Even the ST method can reach 
the best sensitivity of the current 
experiments with a $\rm 0.3 m^3\cdot year$ of exposure at {\kam}.
With the FT method, it it possible to explore the MSSM prediction region  
via SD interactions with a sufficient 
exposure ($\sim~\rm 30 m^3 \times 10 years$). 
When we see any implication of the WIMP-wind, 
a precise study can be performed with the same technology 
by analyzing the shape of the recoil angle distribution.
A prototype WIMP detector with a detection 
volume of 30 $\times$ 30 $\times$ 30 $\rm cm^3$ is now being manufactured. 
Since the fundamental manufacturing technology is already established, 
a large-volume detector ($\rm \sim 1m^3$) 
for underground measurements will soon be available.

\section{Conclusions}
\label{conclusions}
In this paper, we have shown that $\mu$-TPC filled 
with $\rm CF_4$ gas is a promising device for WIMP-wind detection 
via SD interactions.
By the Full-Tracking method with sufficient exposure, it is expected that 
we can not only detect the WIMP-wind, but can also precisely study the 
nature of WIMPs.

\section*{Acknowledgements}
This work is supported by a Grant-in-Aid in Scientific Research of 
the Japan Ministry of Education, Culture, Science, Sports, Technology; 
Ground Research Announcement for Space Utilization promoted by Japan Space
 Forum; and Grant-in-Aid for the 21st Century COE 
''Center for Diversity and Universality in Physics''



\begin{thebibliography}{1}

\bibitem{ref:WMAP Spergel} D.~N.~Spergel {\it et al.,} Astrophys. J. Suppl. 
{\bf 148} (2003) 175
\bibitem{ref:Nelson_review}
H. Nelson, 2003 SLAC Summer Institute, Aug., 2003
(http://www-conf.slac.stanford.edu/ssi/2003); 
{\it Proceedings of the Fourth International Workshop on ``the Identification of Dark Matter'', York, 2002}, edited by N. J. C. Spooner and V. Kudryavtsev, (World Scientific, 2003)
\bibitem{ref:DAMA annual}R. Bernabei {\it et al.,} Phys. Lett. B {\bf 424} (1998) 195; {\bf 450} (1999) 448; {\bf 480} (2000) 23; astro-ph/0307403.
\bibitem{ref:DAMA_aniso}
P.~Belli, R.~Bernabei, A.~Incicchitti, D.~Prosperi, Nuovo Cimento C {\bf 15} (1992) 475; R.~Bernabei, P.~Belli, F. Nozzoli, A. Incicchitti, Eur. Phys. J. C 28 (2003)203.
\bibitem{ref:Buckland_PRL} K.~N.~Buckland, M.~J.~Lehner, G.~E.~Masek, and M.~Mojaver, Phys. Rev. Lett. {\bf 73} (1994) 1067.
\bibitem{ref:APP_CH4}
M.~J.~Lehner, K.~N.~Buckland, and G.~E.~Masek, 
Astroparticle Physics {\bf 8} (1997) 43. 
\bibitem{DRIFT_NIM}
D.~P.~Snowden-Ifft, T. Ohnuki, E.~S.~Rykoff, and C.~J.~Martoff, Nucl. Instrm. Methods A {\bf 498} (2003) 155.
\bibitem{DRIFT_IDM}
D.P.~Snowden-Ifft, C.J. Martoff, and J.M. Burwell, 
Phys. Rev. D {\bf 61} (2000) 101301;
T. Lawson {\it et. al}, in {\it Proceedings of the Fourth International Workshop on ``the Identification of Dark Matter'', York, 2002}, edited by N. J. C. Spooner and V. Kudryavtsev, (World Scientific, 2003), p. 338.
\bibitem{ref:Pikachu}
Y. Shimizu, M. Minowa, H. Sekiya, and Y. Inoue, Nucl. Instrm. Methods A {\bf 496} (2003) 347; H. Sekiya, M. Minowa, Y. Shimizu, Y. Inoue, and W. Suganuma, Phys. Lett. B {\bf 571} (2003) 132
\bibitem{ref:Spergel_WIMP}D. N. Spergel, Phys. Rev. D {\bf 37} (1988) 1353.
\bibitem{ref:Gerbier}J.~Rich and M.~Spiro, Saclay preprint DPhPE 88-04(1988);
G.~Gerbier, J.~Rich, M.~Spiro, and C. Tao {\it Proceedings of the Workshop on ``Particle Astrophysics''}, edited by E.~B.~Norman, (World Scientific, 1989), p. 43.
\bibitem{ref:Masek}G.~Masek, K.~Buckland, and M.~Mojaver,  {\it Proceedings of the Workshop on ``Particle Astrophysics''}, edited by E.~B.~Norman, (World Scientific, 1989), p. 41.
\bibitem{ref:Ellis91}J. Ellis and R. A. Flores, Phys. Lett. B {\bf 263} 
(1991) 259.
\bibitem {ref:SIMPLE}J.~I.~Collar {\it et al.,} Phys. Rev. Lett. {\bf 85} (2000) 3083.
\bibitem{ref:miuchi_APP}
K.~Miuchi {\it et. al}, Astroparticle Physics {\bf 19} (2003) 135; A~.Takeda {\it et. al}, Phys. Lett. B572 (2003) 145.
\bibitem{ref:NIM_CF4_Schmidt}
B. Schmidt and S. Polenz, Nucl. Instrm. Meth. A {\bf 273} (1988) 488.
\bibitem{ref:NUMU}
M.~Avenier {\it et al.,} Nucl. Instrm. Meth. A {\bf 482} (2002) 408
\bibitem{ref:NUMU_DM}
J.L.~Vuilleumier, {\it Proceedings of the Dark Side of the Universe: 
Experimental Efforts and Theoretical Framework, Roma, Italy, 1993,}
edited by R. Bernabei and C. Tao
(World Scientific), p. 281.
\bibitem{ref:Collar_NIM}
J.I.~Collar and Y. Giomataris, Nucl. Instrm. Meth. A {\bf 471} (2001) 254
\bibitem{uPIC:Ochi}
A. Ochi {\it et al.,} Nucl. Instrm. Meth. A {\bf 471} (2001) 264; {\bf 478} (2002) 196.
\bibitem{ref:Nagayoshi_PSD}T.~Nagayoshi {\it et al.,}  Nucl. Instrm. Methods 
A513 (2003) 277.
\bibitem{TPC_PSD:Kubo}
H.~Kubo {\it et al.,} Nucl. Instrm. Methods A513 (2003) 94.
\bibitem{IEEE:Miuchi}
K.~Miuchi {\it et al.,} IEEE Trans. Nucl. Sci., {\bf 50} (2003) 825.
\bibitem{ref:takeda_IEEE}
A~Takeda {\it et al.,} in preparation.
\bibitem{ref:nagayoshi_imaging2003} 
T.~Nagayoshi {\it et al.,} 
Submitted for Nucl. Instrm. Methods A,
 {\it Proceedings of the International Conference on Imaging Techniques 
in  Subatomic Physics, Astrophysics, Medicine, Biology and Industry 
(Imaging2003)", July, 2003}, Stockholm, Sweden.
\bibitem{ref:Miuchi neutron}
K.~Miuchi {\it et al.,} to appear in Nucl. Instrm. Meth. A., physics/0308097
\bibitem{ref:Finland}M. Bouianov {\it et. al.};  T. Nagayoshi {\it et. al.} 
in preparation.
\bibitem{ref:Sharma} A. Sharma, SLAC-Journal-ICFA {\bf 16} (1998) 3.
\bibitem{ref:SRIM}
J. F. Ziegler, J. P. Biersack and U. Littmark,  SRIM - The Stopping and Range of Ions in Matter, Pergamon Press, New York, 1985.
\bibitem{ref:Gilbert}
G.L. Cano, Phys. Rev. {\bf 169} (1968) 277.
\bibitem{ref:DRIFT_CS2}
C. J. Martoff {\it et al.,} Nucl. Instrm. Methods A {\bf 440} (2000) 355.
\bibitem{ref:APP_CF4_Lehner}
M.J.~Lehner, K.N. Buckland, and G.E. Masek, Astroparticle Physics {\bf 8} (1997) 43.
\bibitem{ref:LewinSmith}J.D. Lewin and P.F. Smith, Astroparticle Physics {\bf 6} (1996) 87.
\bibitem{ref:n_at_Kam}A.~Minamino, University of Tokyo (private communication).
\bibitem{ref:uk2000} N.~J.~C.~Spooner {\it et al.,} Phys. Lett. B {\bf 473} (2000) 330. 
\bibitem{SUSY_Ellis} J. Ellis, A. Ferstl, and K.A. Olive,  Phys. Rev. D {\bf 63} (2001) 065016.
\end{thebibliography}
\end{document}